\documentclass[journal=jctcce,manuscript=article]{achemso}
\usepackage{xcolor}
\usepackage{graphicx}
\usepackage{dcolumn}
\usepackage{bm}
\usepackage{soul}
\usepackage[utf8]{inputenc}
\usepackage[T1]{fontenc}
\usepackage{mathptmx}
\usepackage{float}
\usepackage{lipsum}
\usepackage{caption}
\usepackage{subcaption}
\usepackage{amssymb}
\usepackage{mathrsfs}
\usepackage{amsmath}
\usepackage{mathtools}

\author{Juan Pablo Miranda-López}
\affiliation{ Dep. Est. de la Materia, F\'isica T\'ermica y Electr\'onica, Universidad Complutense de Madrid, 28040 Madrid, Spain}
\affiliation{ 
GISC - Grupo Interdisciplinar de Sistemas Complejos 28040 Madrid, Spain
}

\author{Emanuele Locatelli}
\affiliation{Department of Physics and Astronomy,  University  of  Padova, 35131 Padova,  Italy}
\affiliation{INFN, Sezione di Padova, via Marzolo 8, I-35131 Padova, Italy}
\email{emanuele.locatelli@unipd.it}

 \author{Chantal Valeriani}
\affiliation{ Dep. Est. de la Materia, F\'isica T\'ermica y Electr\'onica, Universidad Complutense de Madrid, 28040 Madrid, Spain}
\affiliation{ 
GISC - Grupo Interdisciplinar de Sistemas Complejos 28040 Madrid, Spain
}
\email{cvaleriani@ucm.es}

\title{Self-organized states of solutions of active ring polymers in bulk and under confinement}

\begin{document}


\begin{abstract}
In the presented work we study, by means of numerical simulations, the behaviour of a suspension of 
active ring polymers in the bulk and under lateral confinement.  
When changing the separation between the confining planes and the polymers' density, we detect 
the emergence of a self-organised dynamical state, characterised by the coexistence of slowly diffusing clusters of rotating disks and faster rings  moving in between them.  
This system represents a  peculiar case at the crossing point between polymer, liquid crystals and active matter physics,  where the interplay between activity, topology and confinement leads to a spontaneous segregation of a one component solution. 

\end{abstract}

\maketitle


\section{Introduction}

The study of active filaments and active polymers has attracted significant interest in the last few years\cite{winkler2020physics}. Indeed, filaments brought out-of-equilibrium by the action of fuel-consuming active units are ubiquitous in Nature, ranging from the intra-cellular~\cite{fletcher2010cell,saintillan2018extensile, mahajan2022euchromatin} to the extra-cellular~\cite{loiseau2020active, elgeti2013emergence, chakrabarti2022multiscale} domain, encompassing  unicellular~\cite{balagam2015mechanism, faluweki2022structural, patra2022collective} as well as complex multi-cellular~\cite{deblais2020rheology, deblais2020phase, nguyen2021emergent, patil2023ultrafast} organisms. More recent research has focused on the development of 
artificial systems \cite{ozkan2021collective, zheng2023self}, harbouring promising technological applications.
Among active filaments, the importance of topology has been underappreciated up to now. For passive polymers, topology appears in multiple contexts spanining from DNA supercoiling\cite{van2012dynamics}, to entanglements\cite{Everaers:2004:Science}, cyclisation\cite{haque2020synthesis} or supra-molecular linked materials\cite{hart2021material}. In active filaments, entanglements may be relevant in the context of the observed phase separation in worms\cite{deblais2020phase, patil2023ultrafast}. From a theoretical point of view, different active rings models have been proposed, in different conditions\cite{Active_topoglass_NatComm20, mousavi2019active, philipps2022dynamics, kumar2023local} and all presented a rich phenomenology.\\

Confinement plays a key role in Soft and Active Matter. In the case of polymers, confinement is pivotal for 
the organisation of polymers and biopolymers, as well as for intra-cellular active structures. 
\cite{DNAreview}. Notably, topology and confinement give rise to unique structures, such as the Kinetoplast DNA (or KDNA)\cite{chen1995topology, he2023single}. On top of this, active matter under confinement exhibits a wealth of non-trivial properties\cite{bechinger2016active} and many different systems can be gathered in this category, from micro-swimmers in very narrow or corrugated channels\cite{Bisht2020} to bacteria 
moving through an asymmetric ratchet\cite{Galajda2007,Reichhardt2002, Angelani2011,ten2015can} or in soil\cite{Mattick2002,Henrichsen1983} and epithelial layers\cite{Henkes2020,Morris2019}. For active filaments, confinement has been employed in the modelisation of chromatin dynamics\cite{saintillan2018extensile, mahajan2022euchromatin, liu2021dynamic} with a focus also on micro-phase separation\cite{chubak2022active}. However, the investigation of generic active filamentous systems under confinement has  been fairly neglected, with the exceptions of studies on very short polymeric chains\cite{manna2019emergent, kurzthaler2021geometric,theeyancheri2022migration}.

For these reasons, we are interested  in numerically studying the interplay between activity, topology and confinement in a suspension of active ring polymers. For our purpose, we design activity acting on the polymers as 
polar (or tangential),  
given that this is the active polymer model that displays the richest configurational and dynamical scenario at the single chain level\cite{locatelli2021activity}. In polar active polymers, non-equilibrium phenomena arise due to the interplay between activity and the chain local conformation. We expect that the presence of other rings in the suspension and, possibly, of a steric confinement will introduce  further interplay  enriching the  dynamical scenario.

In our work we study, by means of numerical simulations, a system of  short active rings in bulk or confined between two parallel planes. We show that the interplay between activity, topology and confinement 
gives rise to the emergence of peculiar fluid structures. Under some conditions, we observe the emergence of a self-organised state, characterised by two populations of rings, a cluster of slow rings surrounded by fast rings, that do not belong to said clusters. We demonstrate that this organisation happens at intermediate values of the monomer density and persist in the bulk case. 
We  detect and describe this self-organised state employing a standard clustering algorithm and by looking at single ring conformational and dynamical properties.\\The paper is structured as follows: in Section~\ref{sec:methods}, we report the numerical model and the simulation details, together with the definitions of the observables we are going to consider in our analysis.  We perform our analysis in Section~\ref{sec:structure} by showing how metric properties, such as the gyration radius and the prolateness of individual rings, i.e. their size and shape, averaged over the whole system, show the presence of a strong heterogeneity, which is not present in the extremely diluted case. We detect and characterise the rings' clusters in Section~\ref{sec:clusters}, whose existence is mostly evident under confinement and less in bulk. 
The clustering algorithm allows us to separate rings in clusters from outsiders (i.e. rings that do not belong to any cluster). Finally, in Section~\ref{sec:dynamics}, we use this information to demonstrate that, in the self-organised states, we can effectively define two "populations" with different structural and dynamical properties: rings in clusters being larger and less mobile then  outsiders. Our work  the delicate interplay between activity, topology and confinement in active polymer suspensions.

\section{Model and Methods}
\label{sec:methods}

\subsection{Numerical model}

{We model the active polymer rings in a coarse-grained fashion, employing the well known bead-spring Kremer-Grest model\cite{kremer1990dynamics}. 
A polymer is made of N bonded monomers and 
all monomers interact via a WCA potential \cite{WCA}}
\begin{equation}
   U^{\mathrm{WCA}}=\begin{cases}
4 \epsilon\left[\left(\sigma / r\right)^{12}-\left(\sigma / r\right)^{6}+\frac{1}{4}\right], \qquad &\text{$r < 2^{1 / 6}$} \sigma \\
0, &\text{else} 
\end{cases}
\label{eq:lennardjones}
\end{equation}
{were we set $\epsilon = 50 k_B T$,  $k_B$ being the Boltzmann factor and $T$ being the absolute temperature.
Neighbouring monomers along the ring are held together by the FENE potential} 
\begin{equation}
  U^{\mathrm{FENE}}=\begin{cases}
-0.5 K R_{0}^{2} \ln \left[1-\left(r/ R_{0}\right)^{2}\right], & r \leq R_{0} \\
\infty, &\text{else}
\end{cases}
\label{eq:fene}
\end{equation}
{where we set $K=$30$\epsilon /\sigma^2$=1500$k_B T/\sigma^2$ and $R_0=1.5 \sigma$, being the latter the monomer diameter. This choice is dictated by the fact that we must avoid strand crossings, which can easily appear in the active polymer model employed here\cite{locatelli2021activity}.} Throughout the work, 
we will set  $\sigma$ and the thermal energy $k_B T$
as the units of length and energy, respectively. We also set a unitary mass, so that the characteristic simulation time  $\tau$ is also unitary.

{We introduce polymer's activity  in the form of a tangential self-propulsion along the ring. On each monomer $i$ at position ${\bf r}_{i}$, acts an active force ${\bf f}^{\rm act}_{i}$ with constant magnitude $f^{\rm act}$ and directed at all times along the direction of the vector ${\bf r}_{i+i} - {\bf r}_{i-i}$, parallel to the polymer backbone tangent\cite{bianco2018, foglino2019}.}
{As usual, we express the magnitude of the activity via the adimensional P\'eclet number defined as}
\begin{equation}
\mathrm{Pe}=F_a \sigma/k_B T    
\end{equation}
where $F_a$ is the modulus of the active force. $\mathrm{Pe}$ measures the activity's strength in relation to the thermal noise.

\subsection{Simulation details}

{ We study suspensions of unlinked and unknotted active ring polymers in bulk and confined between two parallel planes.  
A suspension consists of  $M=500$ rings, each one formed by $N=76$ monomers. The rings are evolved according to  Langevin Dynamics, simulated  by means of  the open source package LAMMPS\cite{plimpton1995fast}, with in-house modifications to implement the tangential activity.}
{We integrate the equations of motion using the Velocity Verlet algorithm, with elementary time step $\Delta t = 10^{-4} \tau$ to prevent strand crossings. We employ the standard Langevin thermostat, with friction coefficient $\gamma = 1 \tau^{-1}$.}\\
{As mentioned, we simulate solutions of rings in bulk or under lateral confinement. In the former case, we employ periodic boundary conditions in three directions.  In the latter case, confinement is provided by two perfectly smooth, infinite planes, placed at a distance $h$ and orthogonal to the $z$ direction: the interaction between the flat walls and the monomers is a standard WCA potential, with $\sigma_w=\sigma$ and $\epsilon_w = k_B T$ .}

{We further consider the ring solution at different values of the monomer density $\rho= M \cdot N / V$, where $V$ is the volume of the simulation box. In particular we consider $\rho=$0.05, 0.1, 0.2, 0.3, 0.4, 0.5, thus spanning from the dilute  to the  semi-dilute regime. In the confined case, we consider several values of the separation between the confining plates $h/\sigma=$3, 6, 9, 15, 21, 30. Throughout the text, the bulk case is reported as $h/\sigma=\infty$. }

{Each simulation run starts from a well equilibrated suspension of unknotted, unlinked passive rings. Importantly, we make sure that} chain crossing does not occur {at any time of the simulation} using Kymoknot \cite{tubiana2018kymoknot} {and computing the linking number between any pair of neighbouring rings.}
{After reaching the steady state, we perform production runs of $3-5 \cdot 10^5 \tau$, corresponding to $3-5 \cdot 10^9$ time steps. We save a snapshot of the system every  $\tau_s = 10^{3} \tau$ or $10^7$ time steps. For any given set of parameters, a single independent realisation of the system has been considered.}

{Finally, isolated short active rings in  bulk are expected to undergo a swelling transition at sufficiently large values of $\mathrm{Pe}$. The transition is relatively sharp and the ring conformations are not much affected by the activity at low $\mathrm{Pe}$. Thus, the system is expected to significantly deviate from the passive, homogeneous liquid structure only at high values of the activity\cite{locatelli2021activity}.
Therefore, in this work we will only consider the limit of large activity by setting $\mathrm{Pe}$=10. 
}\\


\subsection{Structural and dynamical analysis}
\subsubsection{Structural analysis}
{In order to characterise the structure of each polymer ring, we compute properties such as  the gyration tensor, the gyration radius and the prolateness. 
The gyration tensor is defined as
\begin{equation}
    G_{\alpha \beta}=\frac{1}{N} \sum_{i=1}^{N}\left(\mathbf{r}_{i, \alpha}-\mathbf{r}_{\mathrm{cm}, \alpha}\right)\left(\mathbf{r}_{i, \beta}-\mathbf{r}_{\mathrm{cm}, \beta}\right)
\end{equation}
where the indices $\alpha$ and $\beta$ runs over the three Cartesian coordinates ($x,y,z$) of the N monomers of each ring.
The gyration radius of the rings can be computed as 
\begin{equation}
    R_g = \sqrt{\lambda_{1}+\lambda_{2}+\lambda_{3}}
\end{equation}
where $\lambda_1,\lambda_2,\lambda_3$ are the three eigenvalues of the gyration tensor. Moreover, we characterize the shape of the rings by computing the prolateness $S$, defined as }
\begin{equation}
    S=\left\langle\frac{\left(3 \lambda_{1}-I\right)\left(3 \lambda_{2}-I\right)\left(3 \lambda_{3}-I\right)}{I^{3}}\right\rangle
\end{equation}
{where $I=\lambda_{1}+\lambda_{2}+\lambda_{3}$. 
Depending on the values of $S$, the polymers' shape can be oblate or prolate: the  shape is oblate (disk-like object) if $S<0$, isotropic (spherical object) for $S\approx0$ or prolate (cigar-like object) for $S>0$. }

\subsubsection{Cluster analysis}
\label{sec:clustalgo}
{In order to} identify {and highlight the self-organised structures that emerge in the polymer suspension, we perform a cluster analysis.} First we compute the centre of mass (CM) of each polymer ring $\mathbf{r}_{\mathrm{cm}} = \frac{1}{N} \sum_{i=1}^{N} \mathbf{r}_{i}$. {We then perform a clustering analysis using the DBSCAN algorithm \cite{DBSCAN}, implemented in the Python library scikit-learn\cite{scikit-learn}, pre-computing the matrix of the mutual distances in order to correctly account for the periodic boundary conditions.} 
DBSCAN allows to deal with a fluctuating number of clusters,  
detects clusters of arbitrary shapes and is robust in detecting outsiders, i.e.  polymer rings not belonging to any cluster. 
We set  the minimal number of elements in a cluster to 5 (arbitrarily) whereas  
the cut-off distance $d_{cut}$, after performing several tests on all systems, was set to {$d_{cut} = 4 \sigma$} 
{as a reasonable common value}.\\ 
{Since we are considering only the centre of mass of the polymer,} we double check the results of the cluster analysis, {assessing the identification of the clusters by visual inspection.}
{Indeed, the when cut-off values $d_{cut}>4$ were employed, clusters that were visibly separated started to merge and outsiders rings, clearly not belonging to any cluster, were instead classified as part of a cluster. Overall this classification resulted less precise.}


{From the cluster analysis, we will  compute  the average number of cluster $N_c$ and the mean cluster fraction $X$, defined by the number of rings that form clusters, divided by total number of rings. Both quantities are averaged over time, when the system reached steady state.}


\subsubsection{Dynamical analysis}
\label{sec:displacement}
{The presence of the self-organised rotating clusters can be highlighted by analysing the clusters' dynamics. } 
{We focus on the clusters' "short" time behaviour, computing} the absolute value of the displacement of the centre of mass of individual polymers in an interval $\tau_{s} = 1000\tau$
\begin{equation}
    \Delta r =\left| \mathbf{r}_{\mathrm{cm}}(t+ \tau_s) - \mathbf{r}_{\mathrm{cm}}(t) \right|
\end{equation}
which correspond to the sampling interval, reported in Section~\ref{sec:methods}.
{Next, we evaluate the distribution of the displacements, distinguishing between rings that were, in both frames, part of a cluster from rings  identified as outsiders. We choose this observable for two reasons. First, while clusters appear to be very long-lived, the individual rings can leave their cluster in the interval $\tau_s$ and, possibly, join other clusters. Further, clusters can merge in between frames. While it is possible, in principle, to keep track of the clusters in time, most rings join or leave a cluster. Thus, the statistics of long time displacements is rather poor. Second, this quantity is already sufficient to highlight the qualitative difference between rings in clusters and outsiders.} 


\section{Results and Discussion}


\subsection{Structural properties in bulk and under confinement}
\label{sec:structure}
\begin{figure}[h!]
    \centering
    \includegraphics[width=0.49\textwidth]{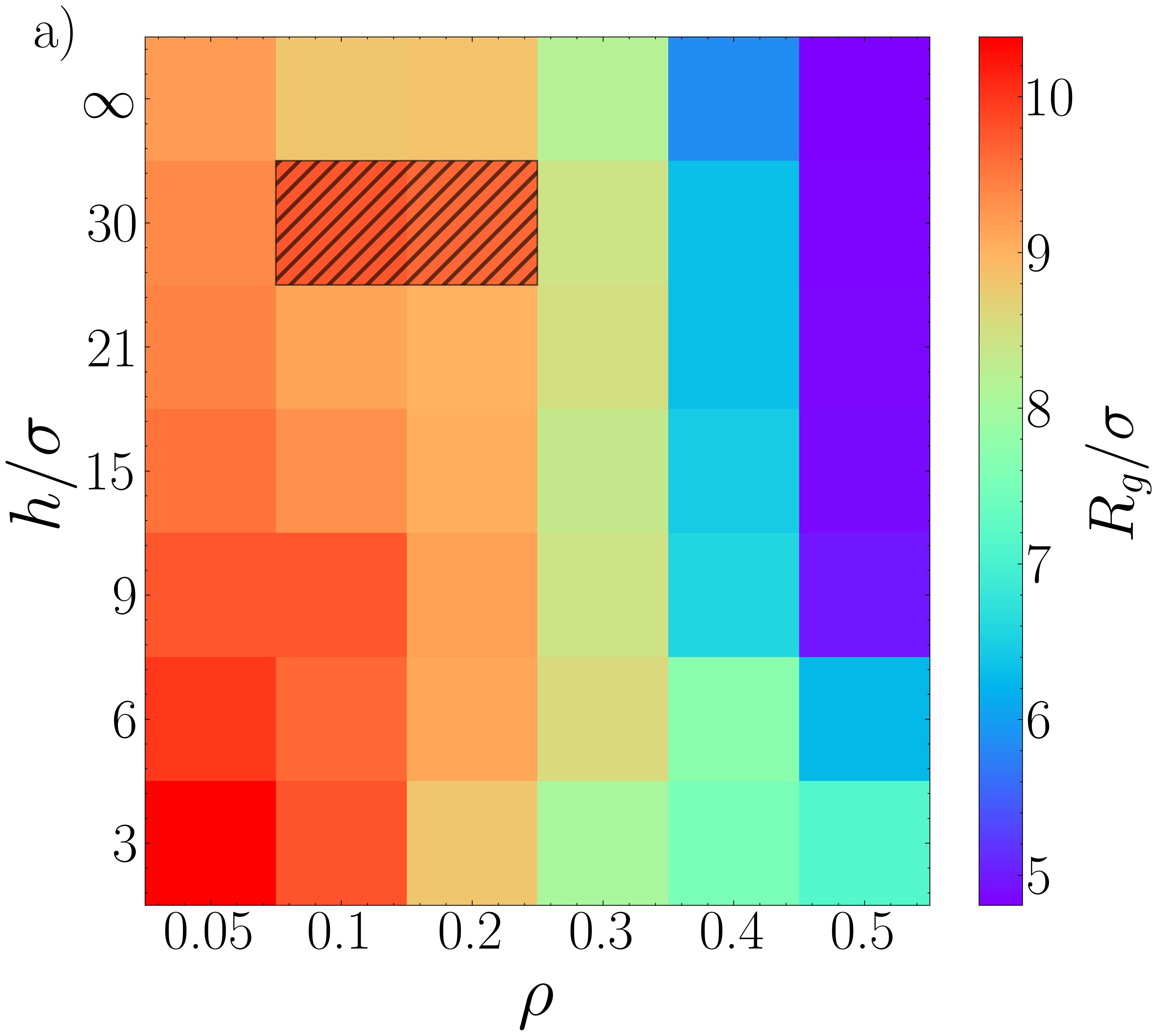}
    \includegraphics[width=0.49\textwidth]{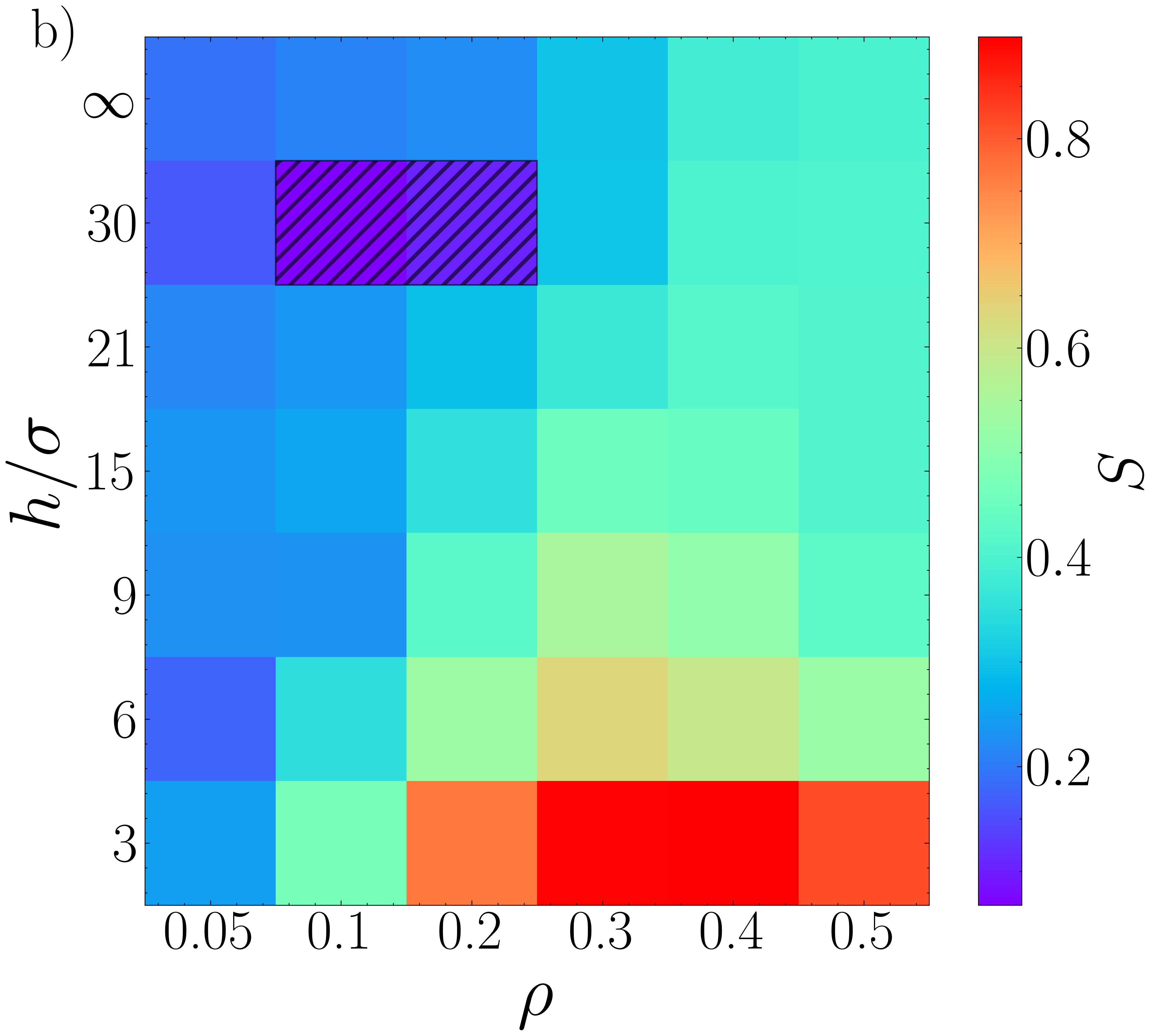}
    \caption{a) Mean gyration radius and b) mean prolateness, for systems of active rings as a function of  density $\rho$ 
    and  lateral confinement $h/\sigma$. 
    The stripes highlight systems with finite size effects that had to be simulated considering larger system sizes.
    }
    \label{fig:metricprops}
\end{figure}

{We first evaluate  structural properties such as   the gyration radius and the prolateness, averaged  over all  rings and over time (in steady state).  In Figure~\ref{fig:metricprops} we report both quantities for different values of  confinement $h/\sigma$  
and for suspensions at different densities.}

{As shown in Figure 1, we observe the emergence of different patterns. Upon increasing the density and, similarly, upon increasing the separation between the confining planes, rings appear to shrink, i.e. the average gyration radius diminishes (Fig.~\ref{fig:metricprops}a). Looking at the prolateness (Fig.~\ref{fig:metricprops}b), upon increasing $\rho$ we observe a change from a roughly spherical shape to a prolate one; instead, upon increasing $h/\sigma$, $S$ diminishes i.e. the rings' shape becomes more spherical. We remark the presence of two special cases, i.e. $\rho=$0.1, 0.2, $h/\sigma=$30, highlighted in both panels of Fig.~\ref{fig:metricprops} by diagonal stripes. For these cases only, simulations showed finite-size effects and we thus performed an additional run with M=2000 rings. The data reported refer to the larger system.

The behaviours described in Figure 1 are  qualitatively different from what is expected in passive systems, and reflect the out-of-equilibrium nature of the system of active polymer rings. In particular, while the trends observed for $R_g$ are not surprising, rings would be expected to become more oblate, rather than prolate, when subject to very high confinement, at least at rather small density\cite{chubak2018ring}. 

However, looking at the distributions of $R_g$ and $S$ across the range of density values, reported in Figure ~\ref{fig:metricdistros}, we discover that their average values, in most cases, does not represent the polymer ring population.  

\begin{figure}[h!]
    \centering
    \includegraphics[width=\textwidth]{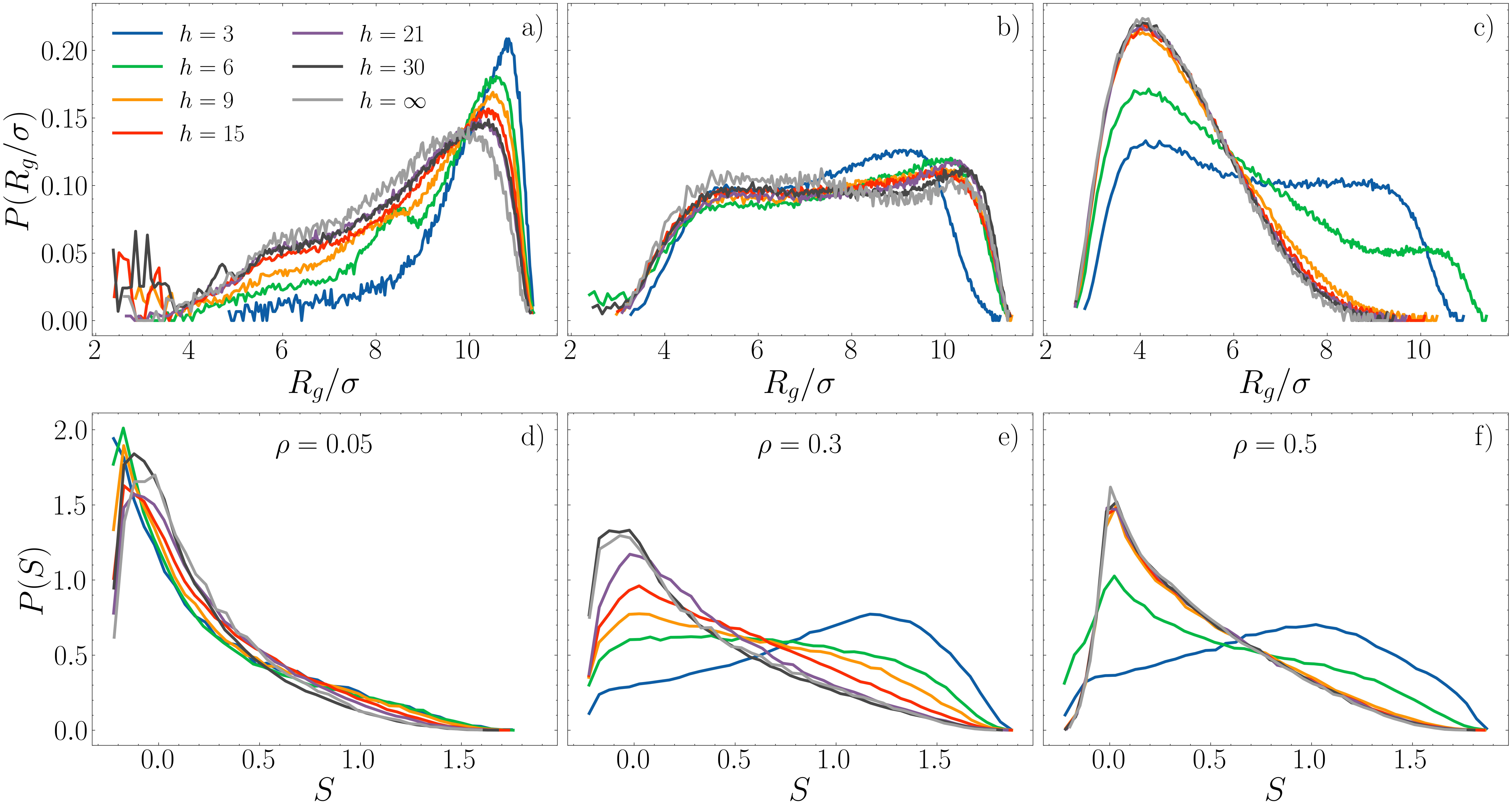}
    \caption{Distribution of $R_g$ and $S$ (top and bottom rows, respectively)  for  densities $\rho=0.1, 0.3, 0.5$ and several values of $h/\sigma$. 
     The bulk system is reported as $h/\sigma=\infty$. 
    }
    \label{fig:metricdistros}
\end{figure}

As shown in the top panels of Fig.~\ref{fig:metricdistros}), the peak of the $R_g$  distribution is displaced, upon increasing the density, to lower values. This explains the average decrease. However, the distribution always spans the same, broad, interval of values. Remarkably, the distributions at $\rho=$0.3 are almost flat: a feature that is, in part, preserved at $\rho=$0.5 for the most confined cases. This implies that extreme conformations, compact or extended, are always detected. 
The decrease observed in this work for a suspension of active polymer rings differs  from the isolated case of either active linear or ring chains\cite{bianco2018, locatelli2021activity}.}\\
As shown in the bottom panels of Fig.~\ref{fig:metricdistros}), 
{at low density, the fact that rings appear spherical emerges as the average of a very asymmetric distribution, presenting a peak at negative values of $S$ and a fat tail extending to extreme prolateness values (see Fig.~ \ref{fig:metricdistros}d). Upon increasing the density, the peak of the distribution of $S$ shifts towards  $S=$0, i.e. the most probable conformation becomes quite spherical. However, the distribution maintains a long tail, which leads to an average positive prolateness (Fig.~\ref{fig:metricdistros}e,f). Finally, notice that the distribution becomes sensitive to the confinement at sufficiently large density: at $h/\sigma$=3, its peak shifts to high positive values, signalling the preference of the rings to assume very elongated conformations. }




{
All in all, the distributions of the gyration radius and of the prolateness suggest the presence of an heterogeneous population of rings with different sizes and shapes.   
}

\subsection{Detecting clusters in the suspensions} 
\label{sec:clusters}
We now  characterise the population heterogeneity emerging from the gyration radius and  prolateness distributions of the rings' shape and size. The distributions, reported in Fig.~\ref{fig:metricdistros}, share similar features in bulk and under confinement. Thus, it is reasonable to suppose that the emerging phenomenology may be common, at least qualitatively. In what follows we will demonstrate  that this is indeed the case. However, one of the feature of the confined system is  that, by visual inspection, the distinction between two different classes of rings clearly emerges. On the contrary, in bulk systems the organisation is visually much less clear; thus, the system in this condition  will not be deepened.

\begin{figure}[h!]
    \centering
    \includegraphics[width=0.45\textwidth]{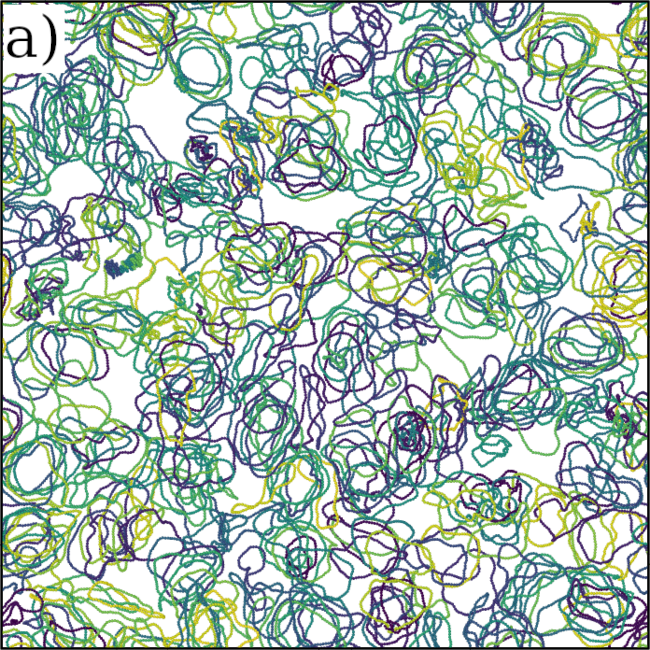}
    \includegraphics[width=0.45\textwidth]{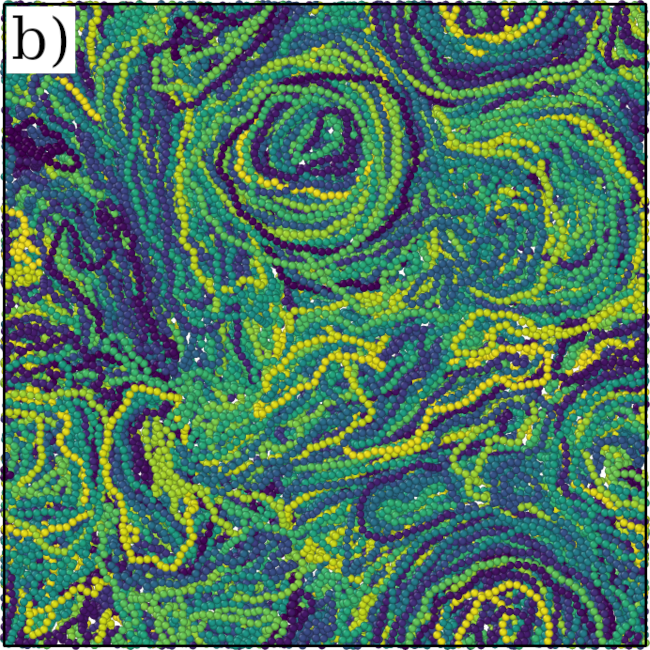}\\
    \includegraphics[width=0.45\textwidth]{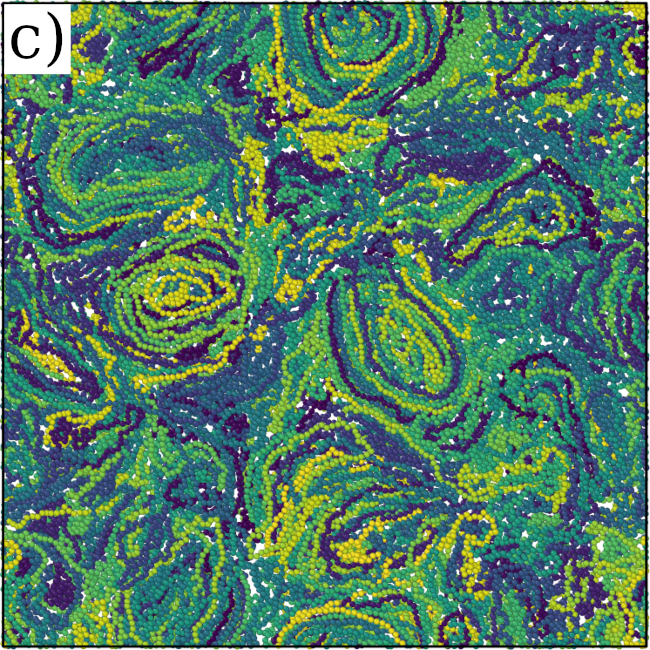}
    \includegraphics[width=0.45\textwidth]{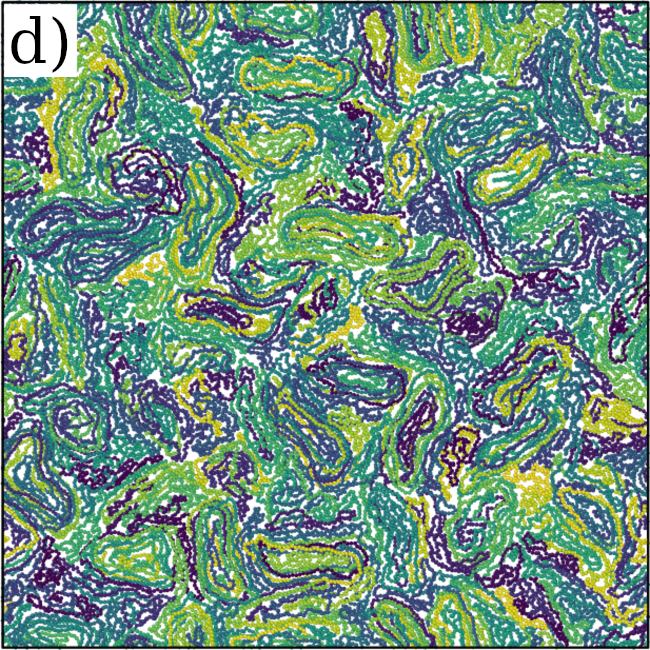}
    \caption{Snapshots of the top view of the active rings suspension (omitting the walls) for different values of $\rho$ and $h/\sigma$. a) $\rho =0.05$, $h/\sigma=15$ dilute case under tight confinement; 
     b) $\rho =0.3$, $h/\sigma=21$ and c) $\rho =0.4$, $h/\sigma=9$ confined cases at intermediate density, where stacked rings are clearly visible; d) $\rho =0.5$, $h/\sigma=3$ most dense and confined case. The colour code follows an arbitrary labelling of the rings and only is used to facilitate the visual distinction of individual chains. }
    \label{fig:snapshot}
\end{figure}

{Figure \ref{fig:snapshot} reports snapshots of active ring suspensions at different values of $\rho$ and $h/\sigma$, seen from the top view, i.e. from the direction perpendicular to the confining planes (the bottom view would be, of course, equivalent). }
{The reported top views show that 
the rings may undergo a process of self-organisation. 
Indeed, rings appear to take peculiar conformations, not common in equilibrium conditions for a passive suspension. 
In steady state, a fraction of the rings is arranged in nest-like structures or clusters. Within these clusters, the rings are rather flat and oblate(see Fig.~\ref{fig:snapshot}b,c) and, as dictated by the activity pattern,  maintain a fairly steady rotation velocity (see video in the Supporting material). Notice that, given that they are visible from the top view, clusters are preferably located  close to the confining walls. The remaining rings moves between clusters  and take a more prolate conformation. Notably, in order to form clusters, rings need a sufficiently high density (see Fig.~\ref{fig:snapshot}a,b) and  space between  confining planes to pile up. When the separation is large enough, clusters  form also across the confining planes (see Fig.~\ref{fig:snapshot}a). One could assume that clustering is favoured at small separations, where it is easier to fill the space between planes with stacked rings, provided  there is enough space. However, the most confined case considered here, $h/\sigma=$3 (Fig.~\ref{fig:snapshot}d), is the one where the least amount of clustering and stacking is detected. In such a case, peculiar conformations of squeezed and elongated rings are favoured. On the other hand, we will later show  the sweet spot where  confinement favours self-organisation (see Sec.~\ref{sec:dynamics}).\\

{In what follows, we will focus on quantitatively characterise   these self-organised states. }
{In order to} identify the self-organised clusters, we {use the DBSCAN} algorithm on the positions of the rings' centres of mass. The chosen cut-off distance, below which two rings are classified as neighbours, is $d_c=4\sigma$, as reported in Section~\ref{sec:clustalgo}.
{Figure \ref{fig:cluster_stuff} represents the cluster fraction $X$ (panel a) and the mean cluster number $N_c$ (panel b) as a function of the monomer density $\rho$ and of the lateral confinement $h/\sigma$.} 
\begin{figure}[h!]
    \centering
    \includegraphics[width=0.45\textwidth]{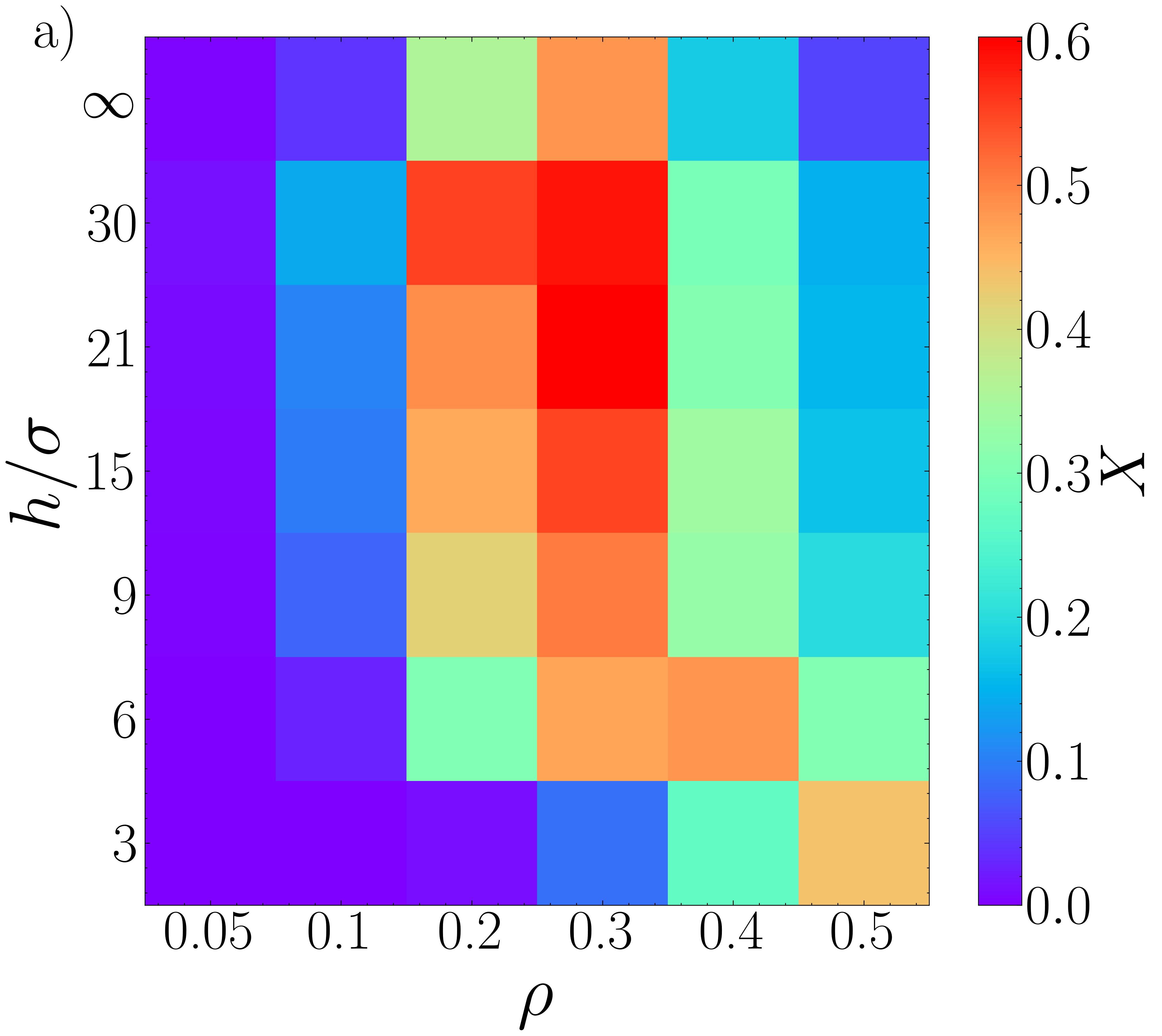}
    \includegraphics[width=0.45\textwidth]{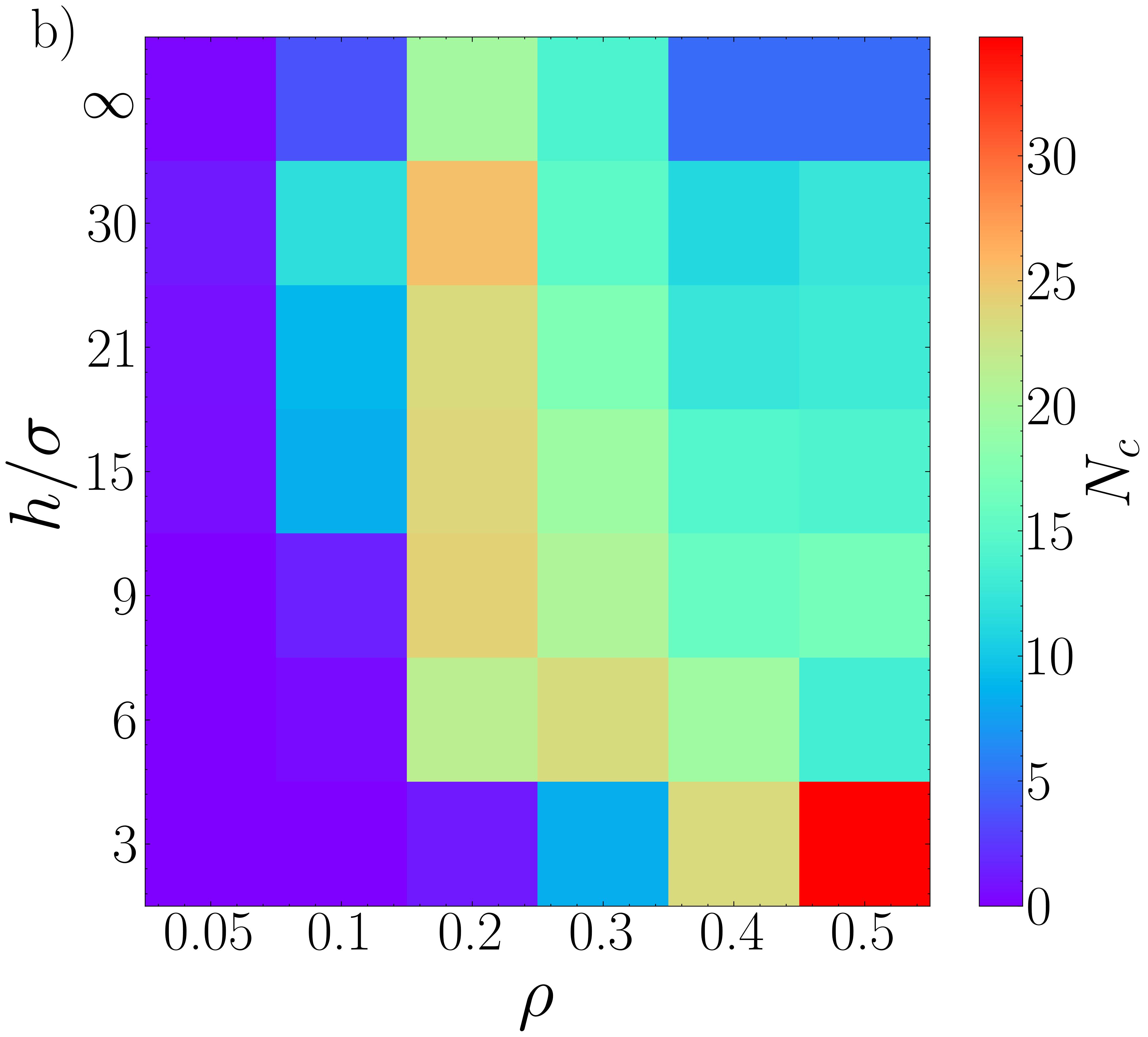}
    \caption{a) Fraction of rings that belong to a cluster $X$, b) mean number of clusters $N_c$ for all   simulated systems.
    }
    \label{fig:cluster_stuff}
\end{figure}
{Clearly, clustering properties strongly depend on both density and confinement. At very high confinement, the algorithm finds clustering only at high density: indeed at $\rho=$0.5 we observe a significant amount of rings classified in clusters. These clusters are, naturally, numerous, as stacks cannot be populated by more than a few rings (see Fig.~\ref{fig:snapshots_clusters}c).} 
{At separations larger than $h/\sigma=$3, the simplest phenomenology happens at very low densities. Here, few clusters are detected, independently of $h/\sigma$. Accordingly, the cluster fraction is low. The interactions between rings are rare and, since there is no attraction, only transient and small clusters can be formed: these are mostly not detected by the algorithm due to our choice of a minimum cluster size of 5 rings. In the rest of the discussion, we will thus not further consider  the case $\rho=0.05$, as no self-organisation is possible.   }


\begin{figure}[h!]
    \centering
    \includegraphics[width=0.45\textwidth]{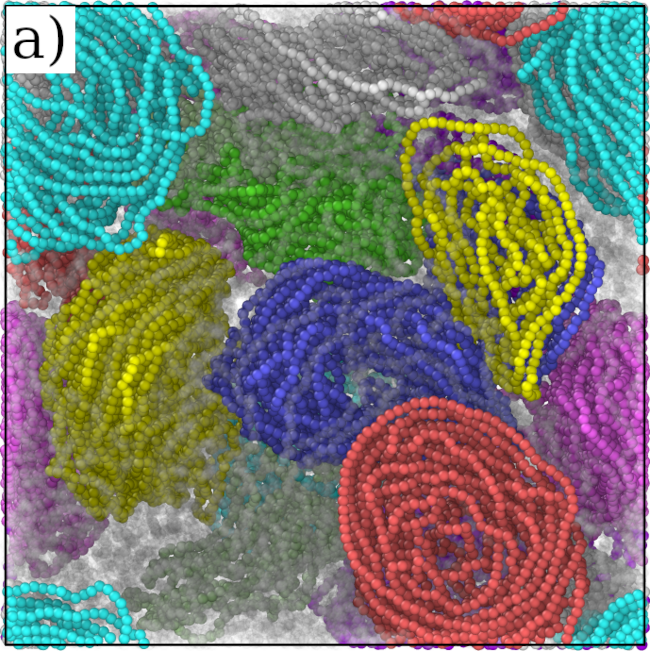} 
    \includegraphics[width=0.45\textwidth]{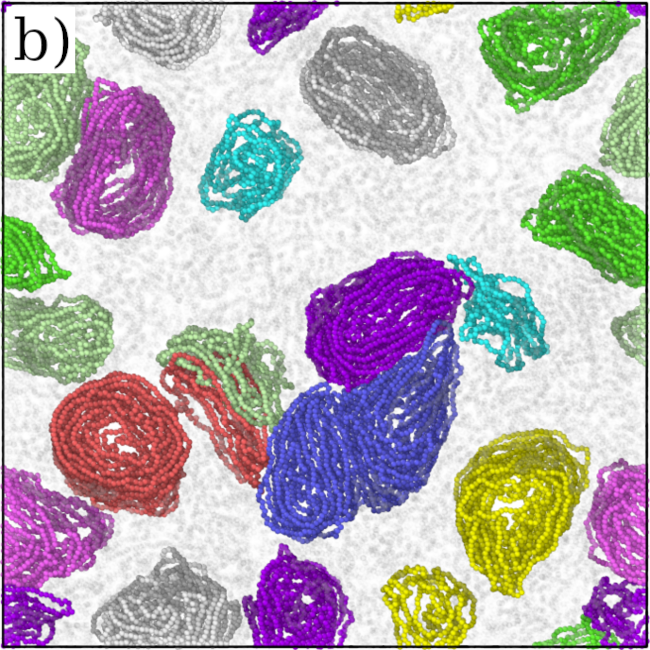} 
    \includegraphics[width=0.45\textwidth]{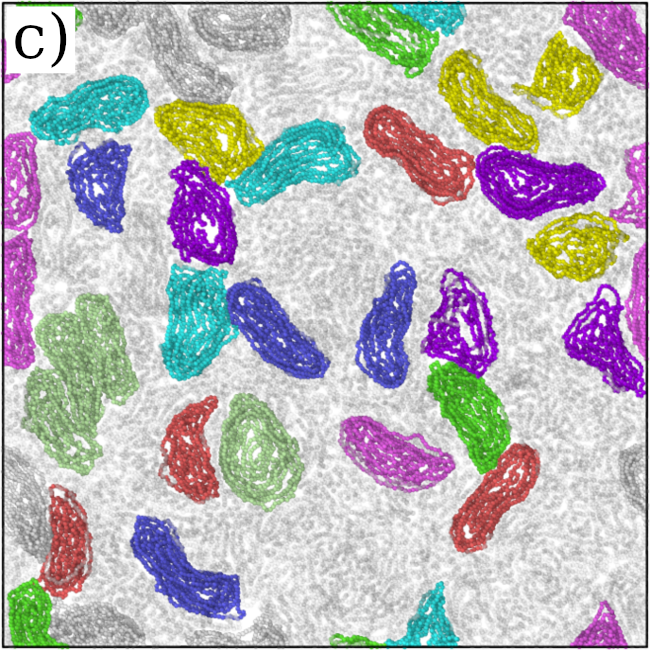} 
    \includegraphics[width=0.45\textwidth]{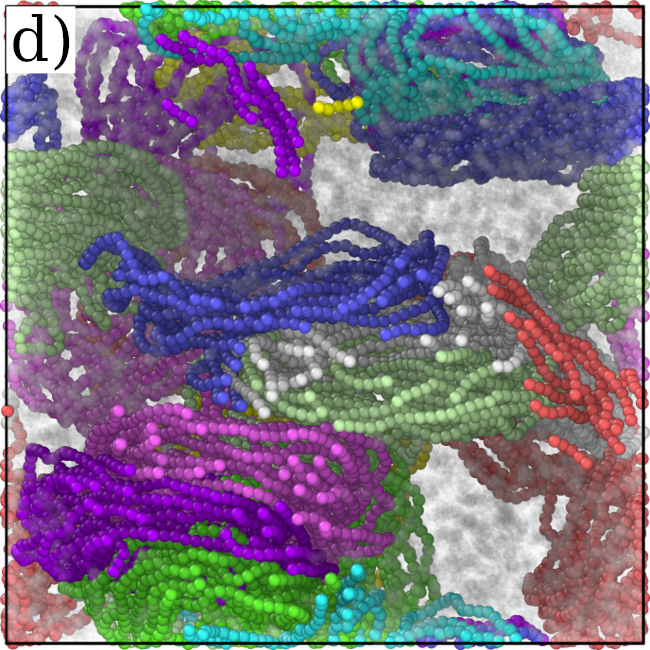} 
    \caption{Snapshots of the active rings suspension, highlighting the ring self-organised clusters, for different values of $\rho$ and $h/\sigma$. a) $\rho = 0.3$, $h=30$. b) $\rho = 0.4$, $h=6$. c) $\rho = 0.5$, $h=3$. c) $\rho = 0.2$ in the bulk. The systems under confinement are seen from the top, as in Fig.~\ref{fig:snapshot}. The polymers labelled as outsiders are shown as transparent in order to highlight the clusters; different colours mark different clusters. }
    \label{fig:snapshots_clusters}
\end{figure}

Upon increasing the density, for $h/\sigma>$3, we notice a sharp growth in the cluster fraction, as well as in the number of clusters. These quantities reach a maximum value  around $\rho=0.2-0.3$ and, intriguingly, at large plane-plane separations (see Fig.~\ref{fig:snapshots_clusters}a). Figure \ref{fig:snapshots_clusters}d shows that  such stacks are present also in the bulk, where  there is no wall to "nucleate" the clusters: thus, their orientation is random and this makes it difficult to detect them without a suitable algorithm.
{Upon further increasing the density, the cluster fraction and the number of clusters now diminish. Here the cluster analysis underlines its limitations: the system is dense and  rings tend to assume more compact configurations, as noticed in Fig~\ref{fig:metricdistros}. Thus, one might   need to complement the cluster analysis with tools   to have a way to characterise systems where clusters form from systems where clusters are not present. }



\subsection{Structure and dynamics of rings inside and outside of the clusters}
\label{sec:dynamics}
{From the snapshots reported in Fig.~\ref{fig:snapshots_clusters}, it intuitively emerges  that a straightforward way to characterise rings in clusters with respect to outsiders is to highlight the difference in shape and size between the two groups. For this reason, we compute again the distributions of $R_g$ and $S$, separating the rings belonging to  clusters from the outsiders.}

\begin{figure}[h!]
    \centering
    \includegraphics[width=\textwidth]{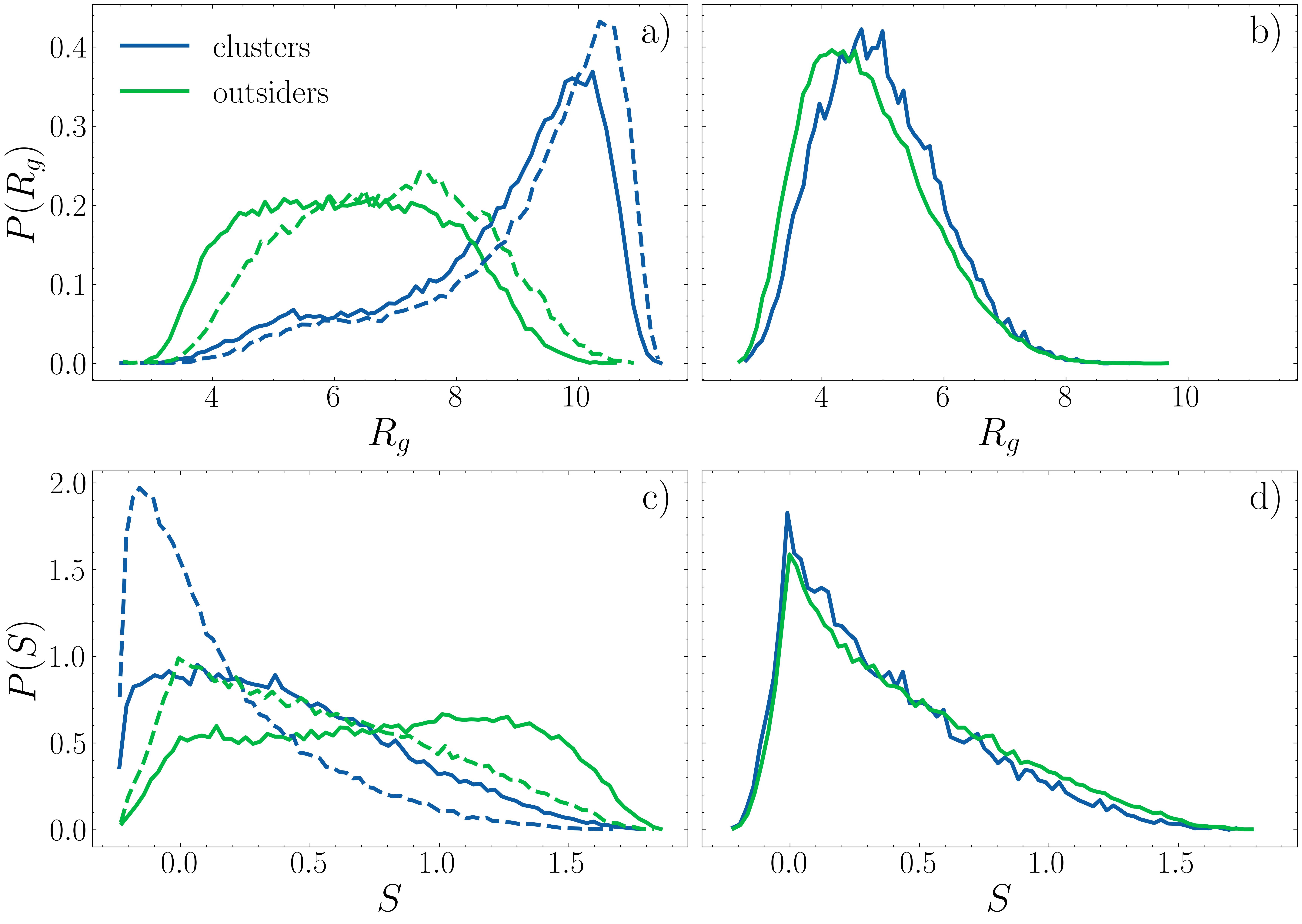}

    \caption{Probability distribution functions of $R_g$ (panels a, b) and $S$ (panels c, d) for rings detected as part of clusters (blue) or outsiders (green). Panels a, c report $\rho = 0.4$, $h/\sigma=6$ (full lines) $\rho = 0.3$, $h/\sigma=30$ (dashed lines); panles b, d report $\rho=$0.5, $h/\sigma=$21. 
    }
    \label{fig:cluster_PDF}
\end{figure}

Few examples are reported  in Fig.~\ref{fig:cluster_PDF}. {When clusters are present, the distributions are markedly different, as in Fig.~\ref{fig:cluster_PDF} a,c), reporting data for $\rho=$0.4, $h/\sigma=$6 (full lines) and $\rho = 0.3$, $h/\sigma=30$ (dashed lines).  

On the one side, rings that take part in the clusters tend to be expanded and oblate, with a distribution that usually has a peak at high values of $R_g$. On the contrary, outsiders rings tend to be prolate and present a more heterogeneous distribution of sizes, peaked at smaller values of $R_g$. 
On the other side, when clusters are not present, the distributions are very similar, as in Fig.~\ref{fig:cluster_PDF} b,d), reporting data for $\rho=$0.5, $h/\sigma=$21. }  This analysis highlights the fact that clusters, detected at large values of $\rho$, are due to local density fluctuations and not to self-organization.

\begin{figure}[h!]
    \centering
    \includegraphics[width=\textwidth]{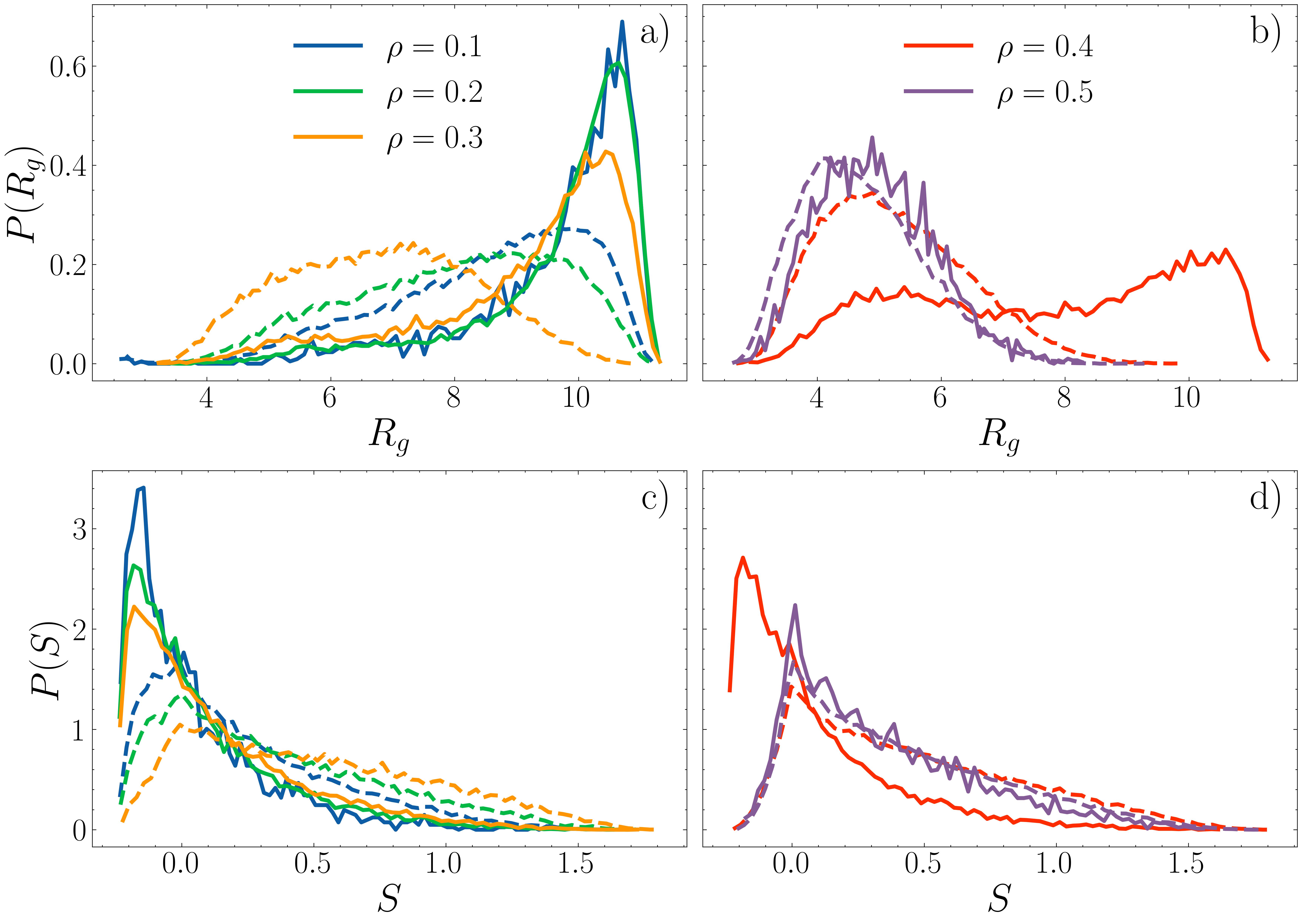}
    \caption{Probability distribution functions of $R_g$ (panels a,b) and $S$ (panels c,d) for active rings in the bulk at different values of $\rho$, detected as part of clusters (full lines) or outsiders (dashed lines).} 
    \label{fig:cluster_bulk}
\end{figure}

{Finally, it is intriguing that  clusters remains  even if  confinement is relatively small, i.e. at large values of $h/\sigma$. As highlighted by Fig.~\ref{fig:snapshots_clusters}d), proper clusters can be also detected in bulk.  Figure~\ref{fig:cluster_bulk} reports the distributions of $R_g$ and $S$ for rings classified as belonging to a cluster or as outsiders for systems in the bulk at different values of $\rho$. The same phenomenology observed under confinement emerges also here, confirming that clusters of disk-like rings form in the bulk within the density interval 0.05$< \rho <$ 0.5.}\\ 

{We employ a similar strategy to investigate the dynamics. We consider the dynamical properties of rings belonging to a clusters and outsiders, separately,} 
and  compute, for both groups, the distribution of  displacements of  individual rings (as detailed in Section~\ref{sec:displacement}). 

\begin{figure}[h!]
     \includegraphics[width=\textwidth]{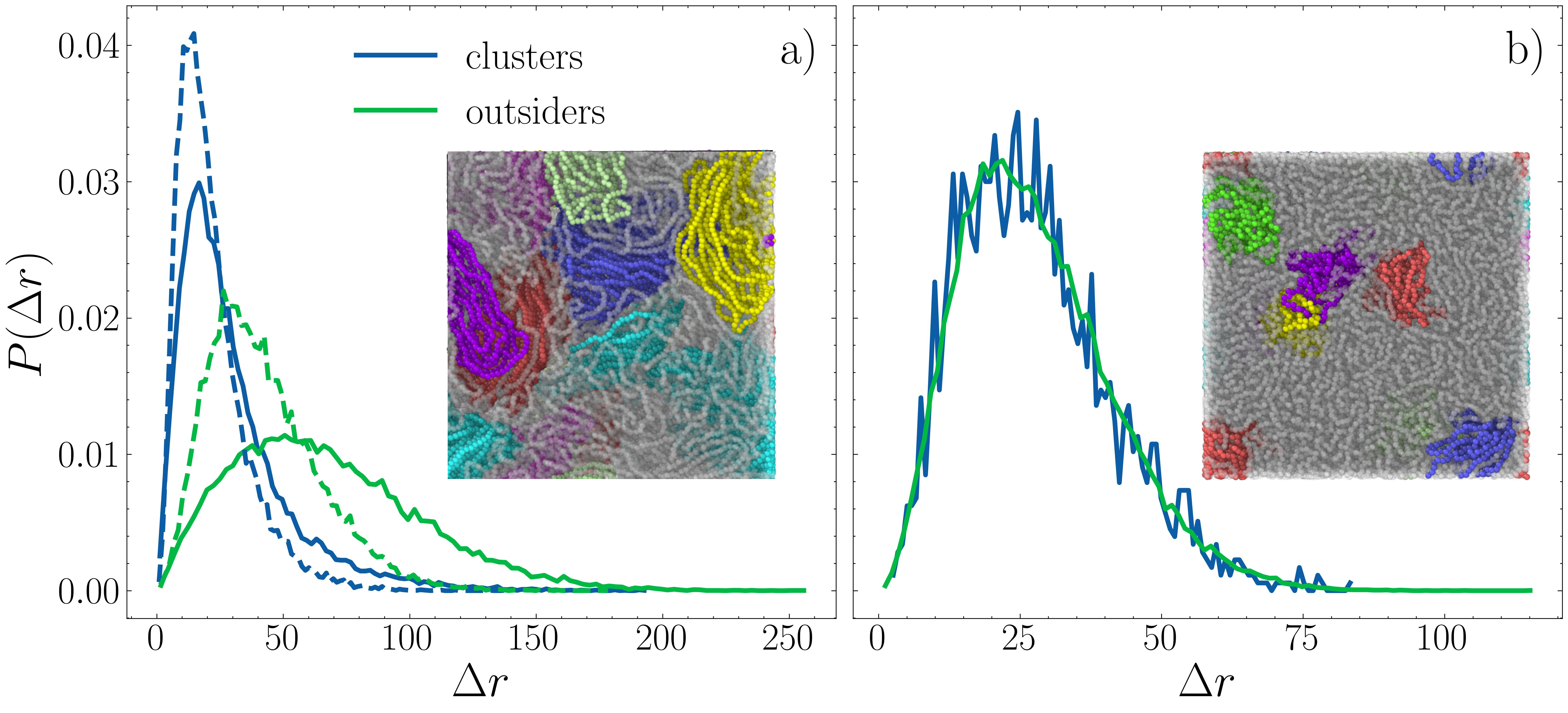}
    
	\caption{ Probability distribution of the displacements $P(\Delta r)$ for rings belonging to clusters (blue line) and outsiders (green line). a) $\rho = 0.4$, $h/\sigma=6$ (full line, snapshot in inset) $\rho = 0.3$, $h/\sigma=30$ (dashed line) b) $\rho = 0.5$, $h/\sigma=21$ (snapshot in inset). In the snapshots, outsiders are shown in grey, and the other colours highlight different clusters.
    }
  \label{fig:probability_displacements}
	\end{figure}
 
Figure~\ref{fig:probability_displacements} shows examples of such dynamical analysis. 
{We  observe the same phenomenon found upon looking at the distributions of the rings' size and shape: the distribution of the rings' displacements  inside  the clusters and of the  outsiders is markedly different (Fig.~\ref{fig:probability_displacements}a) when self-organisation is present. Instead, when no self-organisation emerges, all  rings have the same displacement distribution (Fig.~\ref{fig:probability_displacements}b).} 

¡
{In order to establish a more quantitative comparison between all  systems, we extract, }from the distributions, the most probable value of the displacement, for rings inside clusters and outsiders, $\Delta r_C$ and $\Delta r_O$, respectively. 
\begin{figure}[h!]
    \centering
    \includegraphics[width=0.49\textwidth]{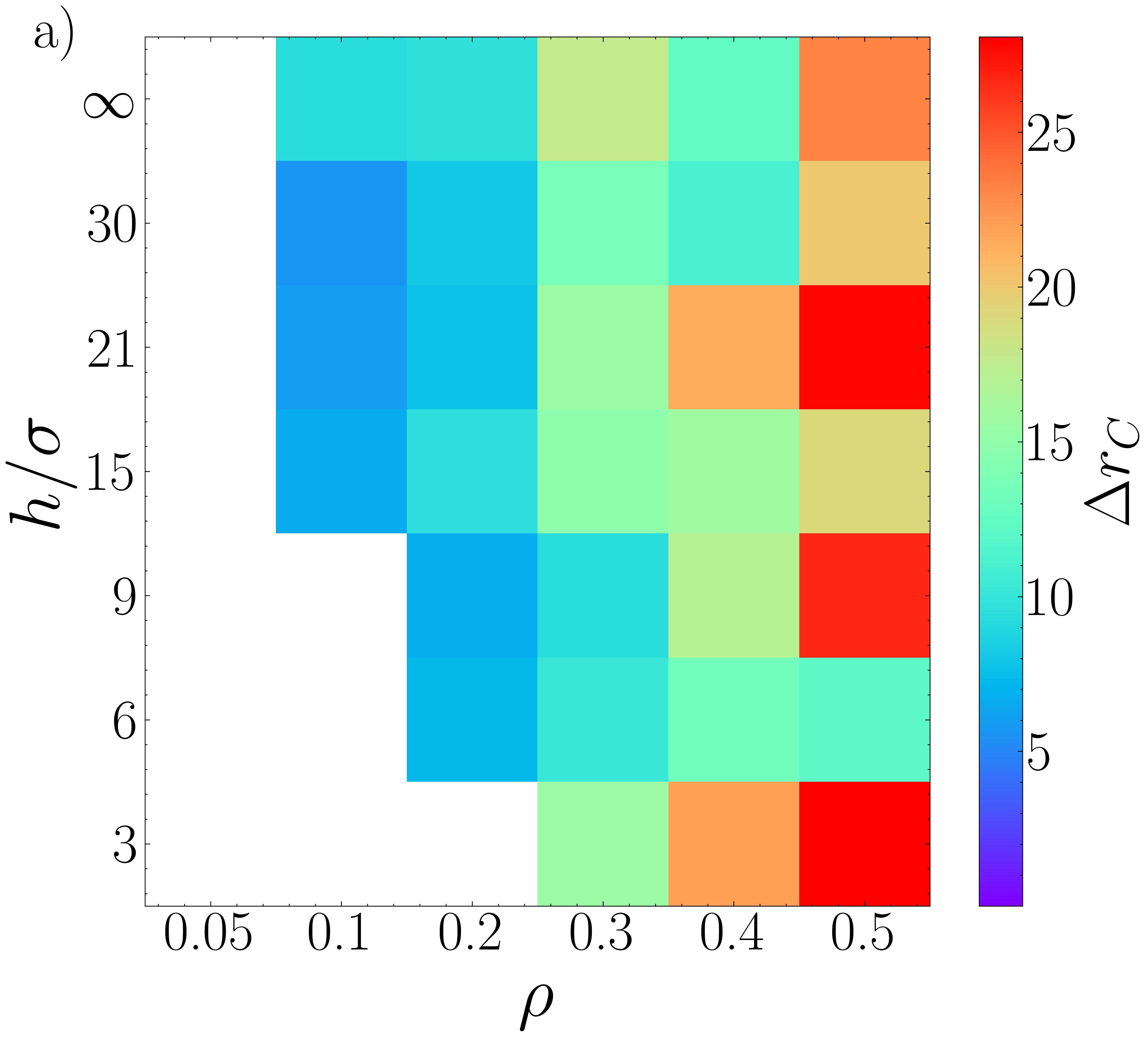}
    \includegraphics[width=0.49\textwidth]{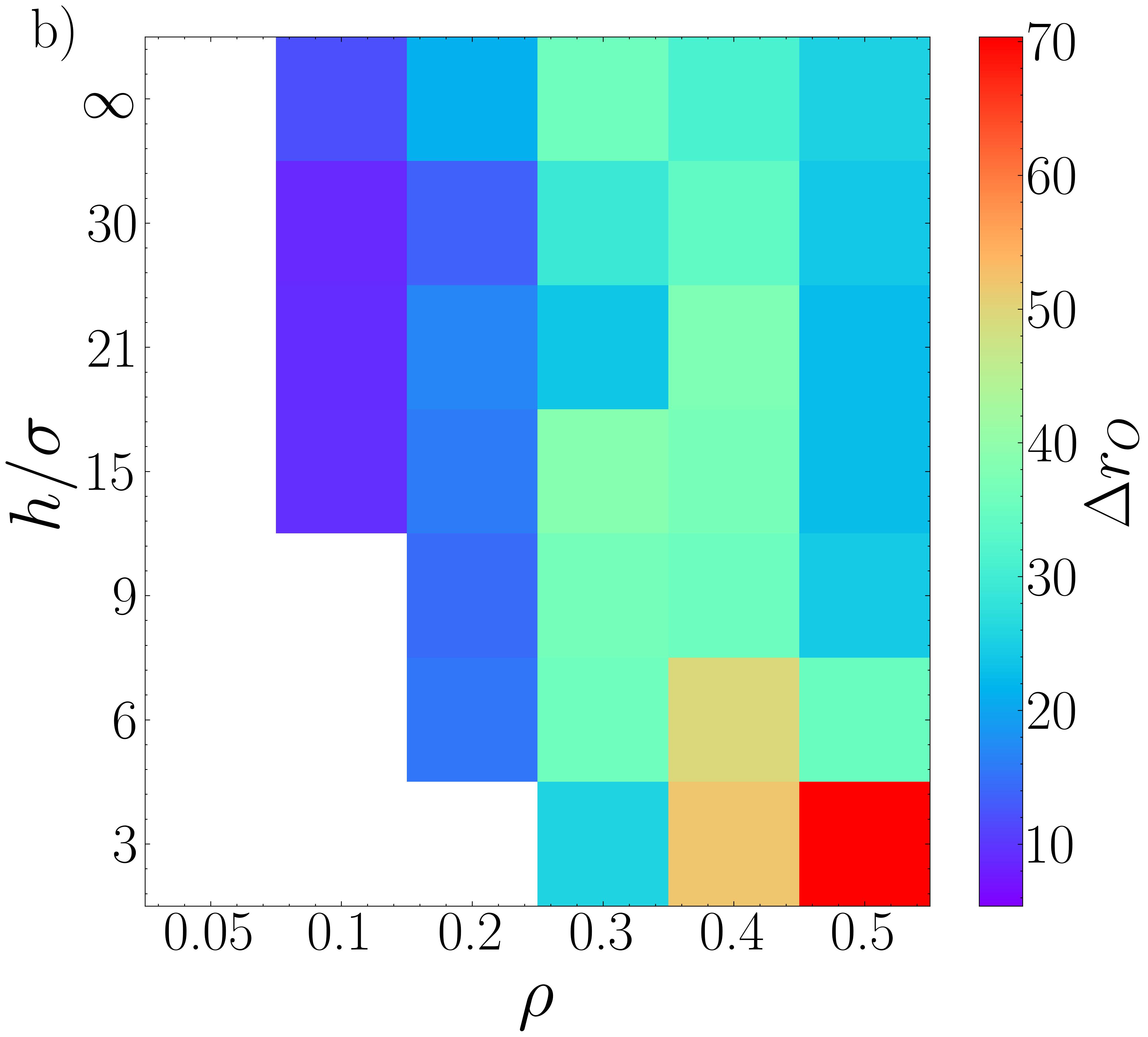}
    \caption{Most probable values of the displacements for (a) rings identified as part of aggregates, (b) ring identified as outsiders. The blank  spaces highlight the system where the number of the identified clusters was negligible.  
    }
    \label{fig:displacement_map}
\end{figure}
We report the results in Fig.~\ref{fig:displacement_map}, 
that can be understood when 
taking into account Fig.~\ref{fig:cluster_stuff}: it is not meaningful to consider systems where there are no clusters. Thus, we leave  blank spaces  and  focus on the rest of the phase diagram. 
By looking at the range of values reported in the two colorbars in Fig.~\ref{fig:displacement_map}a) and b),  rings in} clusters are less mobile than the outsiders. {Counter intuitively, the mobility for rings in clusters increases upon increasing  density: one would expect the opposite for passive systems. The presence of the polar activity introduces a strong interplay between the polymer conformation and dynamics\cite{bianco2018}; rings are more compact at high density and this enhances their mobility with respect to the swollen case. However, clusters at high density are qualitatively different from the ones at low density, as shown above; thus a direct comparison is not straightforward and should be avoided.}\\

\begin{figure}[h!]
    \centering
    \includegraphics[width=0.49\textwidth]{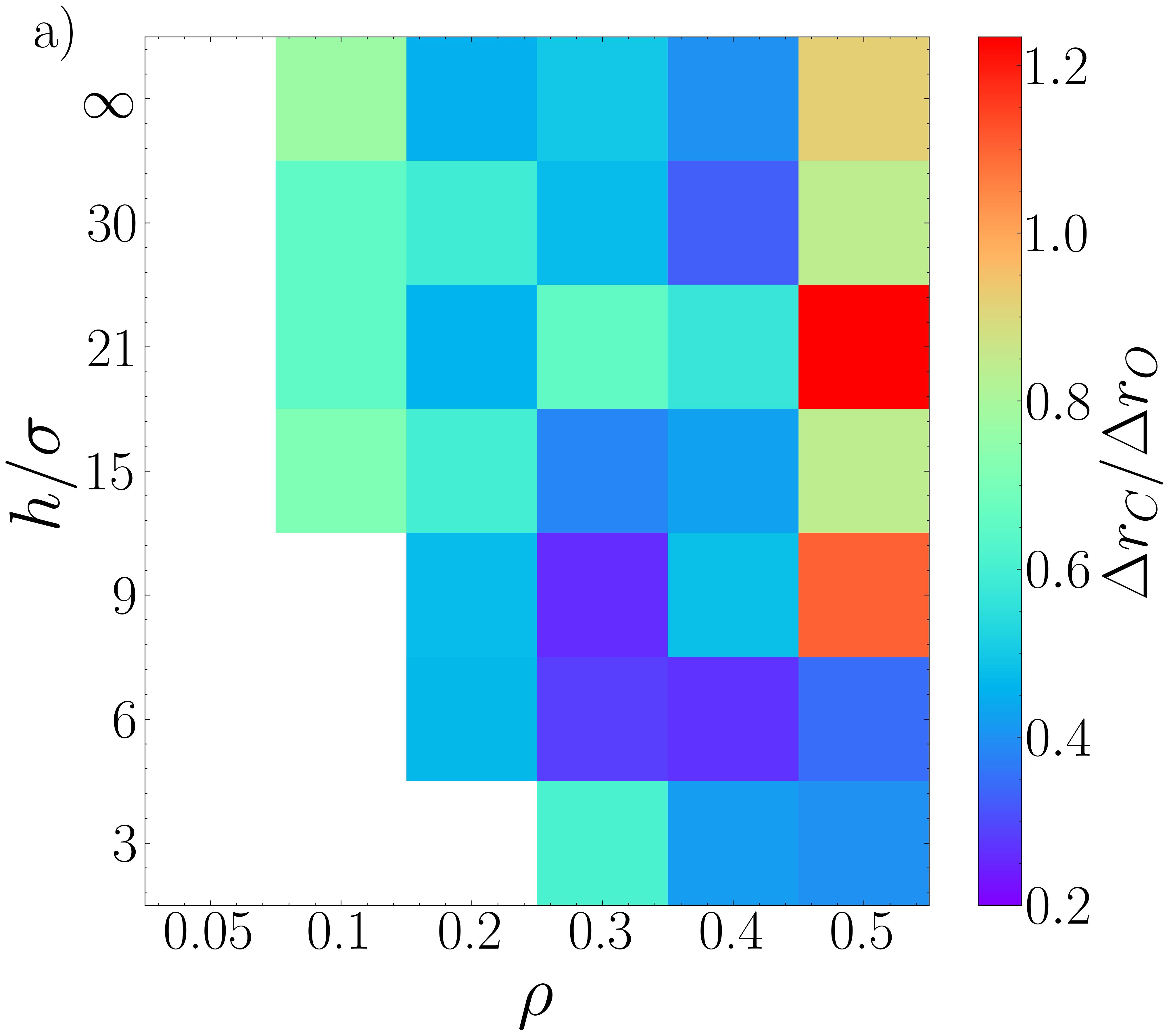}
    \includegraphics[width=0.49\textwidth]{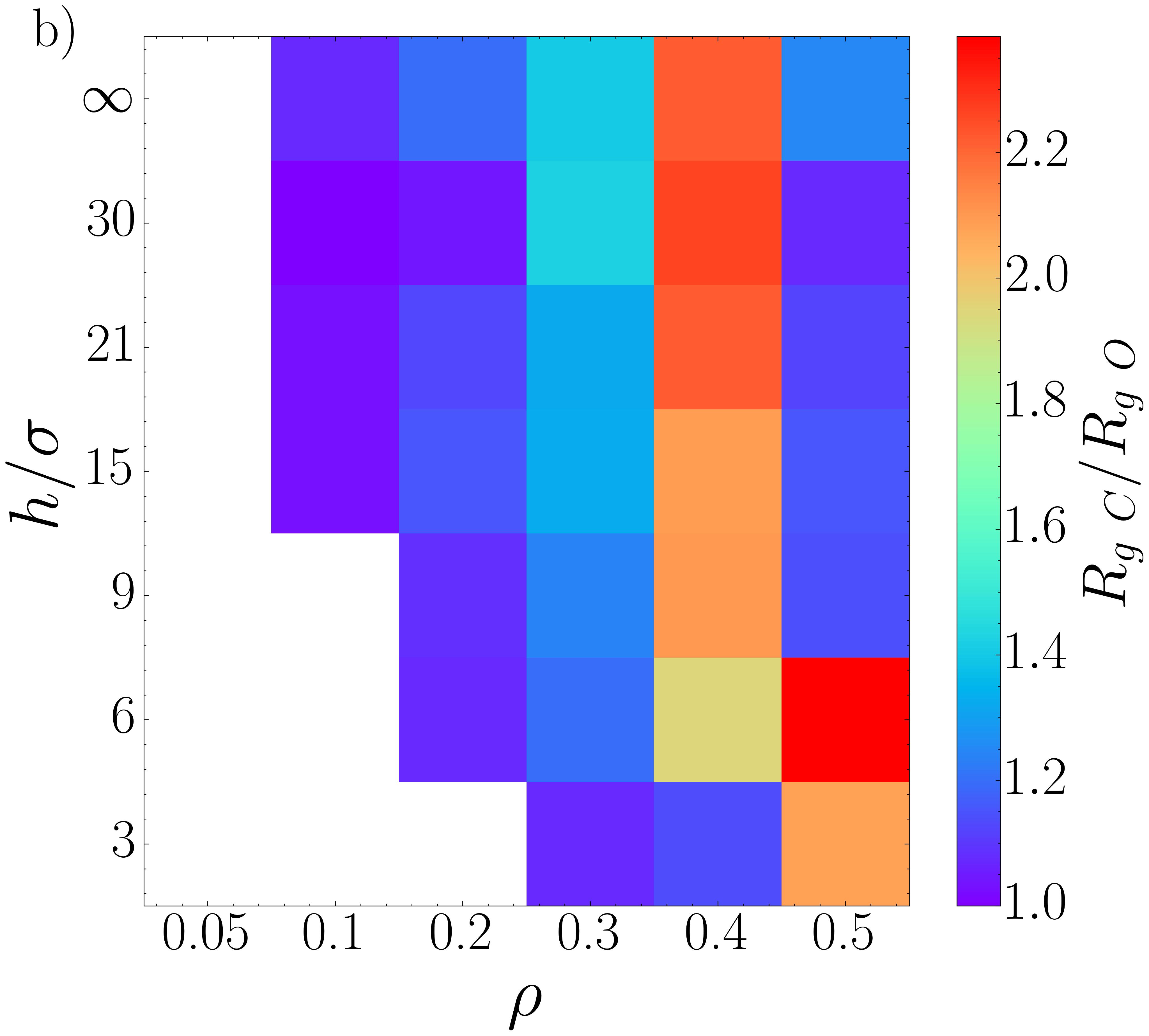}
    \caption {Ratio between the most probable values of a) the displacements b) $R_g$, for rings in clusters and outsiders particles a function of the density $\rho$ and the separation between the confining planes $h\/sigma$. The white colour highlights the system where the number of the identified clusters is negligible.
    }
    \label{fig:displacement_map_2}
\end{figure}

Finally, as a way to quantify the degree of heterogeneity within the two ring populations, we consider the ratio between the most probable values of the observables computed so far, $R_g$ and $\Delta r$, inside and outside the clusters, indicated with subscripts $C$ and $O$, respectively. We report the result in Fig.~\ref{fig:displacement_map_2}. Looking at the the displacements (Fig.~\ref{fig:displacement_map_2}a), systems with rings of different mobility will be characterised by $\Delta r_C/\Delta r_O \ll 1$, otherwise $\Delta r_C/\Delta r_O \approx 1$. Instead, for the gyration radius (Fig.~\ref{fig:displacement_map_2}b), clusters should be characterised by $R_{g,C}/R_{g,O} > 1$, or  $R_{g,C}/R_{g,O} \approx 1$ otherwise.\\We discover, qualitatively, a correspondence in the two panels of Fig.~\ref{fig:displacement_map_2}: systems where clusters form are also  the systems where the size difference is more prominent. Highlighting $\rho=0.5$, $h/\sigma=$3 and $\rho=0.5$, $h\sigma=$6, 
confinement  favours self-organisation when the system becomes denser. 

\section{Conclusions}

In our work, we have explored the conformation and dynamics of  short active polymer rings in bulk and under confinement. We found   quite a large   $\rho$-$h/\sigma$ parameter space where  rings  self-organised in clusters. The cluster formation happened close to the walls under confinement, but also appear in bulk (even though less prominently). One can ascribe the origin of the clusters to the combination of the self-propulsion and of the molecular fingerprint of the rings. Isolated short active rings tend to assume a swollen, disk-like configuration; this was associated to an effective persistence length and to an effective "semi-flexibility"\cite{locatelli2021activity}. In passive systems, semi-flexible rings have been reported to form clusters at high density\cite{poier2016anisotropic}. However, there is a more delicate interplay here to consider. On the one hand, the presence of other active rings perturbs the disk-like conformation observed at infinite dilution, that results from the interplay between  active forces,  elasticity of the backbone and  fixed topology. The activity is, in turn, connected to the polymer conformation; when the latter is perturbed, the action of  active forces give rise to complex configurational rearrangements. Thus, the emergence of clusters in bulk is not trivial, but  rather unexpected and, intuitively, the self-organisation scenario is bound to fail at sufficiently high density, when the inter-chain contacts become dominant.\\  
In addition to this, when considering the system under confinement, the molecular fingerprint of the rings is even more subtle. Passive ring polymer solutions under confinement have been shown to display an inhomogeneous density profile at relatively low density, with a marked tendency to accumulate to walls; even in the passive case, oblate configurations are enhanced close to a wall\cite{chubak2018ring}. Further,  active agents interacting with walls display very distinct emergent patterns, which would not be present in equilibrium\cite{bechinger2016active}; distinctively, they accumulate against walls\cite{Galajda2007, bechinger2016active}. Thus, the emergence of the stacks close to the walls becomes favourable from multiple perspectives: rings can remain in their most favourable configuration, which is indeed enhanced by the presence of the walls and wall accumulation is favoured, as rings can be packed more efficiently. Thus, when $h/\sigma$ is large, rings have two mechanisms to stack: the "bulk" one, i.e. they pile up among themselves in a random orientation and the confined one, i.e. they pile up perpendicularly to the confining wall. This rationalises the excess clustering and self-organisation observed even at quite large separation when, looking at the whole system, one sees little difference between bulk and confined distributions.\\Our computational study demonstrates the possibility of realising self-organising active fluids with polymers, where the peculiar "two population" fluid emerges thanks to the deformability of the rings, that can switch between squeezed, elongated and swollen, disk-like configurations reversibly, without changing their topology. This system indeed shows features, typical of a liquid crystal, as the stacks may be seen as local regions of nematic order; we will investigate these features in more details in the future. Further, it will be interesting to understand how to enhance or suppress the self-organisation; the latter property may be relevant to biological systems such as malaria sporozites\cite{patra2022collective} and may be exploited  as a mechanism to stop cells' development 
in a more focused and environmentally sustainable fashion.

 
\begin{acknowledgement}

J.P. Miranda acknowledges support by a STSM Grant from COST Action CA17139 (eutopia.unitn.eu) funded by COST (www.cost.eu). E. Locatelli acknowledges support from the MIUR grant Rita Levi Montalcini and from the HPC-Europa3 program. C.V acknowledges funding from MINECO grants C.V. acknowledges fundings  EUR2021-122001, PID2019-105343GB-
I00, IHRC22/00002 and 
PID2022-140407NB-C21 from MINECO. The computational results presented have been achieved using the Vienna Scientific Cluster (VSC) and the Barcelona Supercomputing Center (BSC-Marenostrum); CloudVeneto is also acknowledged for the use of computing and storage facilities.
\end{acknowledgement}


\providecommand{\latin}[1]{#1}
\makeatletter
\providecommand{\doi}
  {\begingroup\let\do\@makeother\dospecials
  \catcode`\{=1 \catcode`\}=2 \doi@aux}
\providecommand{\doi@aux}[1]{\endgroup\texttt{#1}}
\makeatother
\providecommand*\mcitethebibliography{\thebibliography}
\csname @ifundefined\endcsname{endmcitethebibliography}
  {\let\endmcitethebibliography\endthebibliography}{}

\end{document}